\documentstyle[epsfig]{jaa}
\begin{document}

\noindent
\title[ Cosmic Microwave
 Background Temperature at 1280~MHz]
{A measurement of  the cosmic microwave
 background temperature at 1280~MHz}

\author[ A. Raghunathan \&   Ravi Subrahmanyan]
{ A. Raghunathan$^{1}$ \& Ravi Subrahmanyan$^{1,2}$ \\
$^{1}$Raman Research Institute, C.V. Raman Avenue, 
Bangalore - 560 080, India \\
$^{2}$Australia Telescope National facility, CSIRO, 
Locked bag 194, Narrabri, NSW, Australia \\
}

\pubyear{2000}
\volume{0}
\pagerange{\pageref{firstpage}--\pageref{lastpage}}
\setcounter{page}{0}
\date{Received 2000 Xxx YY; accepted 2000 Xxx YY}
\maketitle
\label{firstpage}
\begin{abstract}
The absolute temperature of the cosmic microwave 
background  (CMB) has been  measured
at a frequency of 1280~MHz. The observation was made with a 
modified version of the L-band receiver used in the Giant Metre
wavelength Radio Telescope (GMRT): the feed horn was replaced
by a corrugated plate and the receiver was placed on the ground,
directed at zenith, and shielded from ground radiation by an 
aluminium screen with corrugated edges. Novel techniques have been adopted 
for (1) reducing and cancelling unwanted contributions to the
system temperature of the receiver and (2) calibrating the contributions
from the feed assembly and receiver.  The thermodynamic temperature of the 
CMB is estimated to be ${3.45\pm 0.78}~$K.
\end{abstract}


\section{Introduction}
The spectrum of the cosmic microwave background (CMB) has been measured
by the $COBE-FIRAS$ experiment (Fixen et al. 1996) to be very closely 
Plankian in form over the frequency range 70-640~GHz; the best-fit
thermodynamic temperature of the radiation was determined to be 
$2.728 \pm 0.004$~K over this frequency range.  This is consistent with
the expectation --- within the standard big-bang cosmology --- that the
relict radiation was thermalized via free-free and radiative Compton
scattering at redshifts $z > z_{th} \sim 7.5 \times 10^{6}$
(Burigana, Danese \& De Zotti 1991).

The damping of small wavelength (large $k$-mode) sub-horizon scale
baryon adiabatic perturbations  --- as
a result of photon diffusion --- may be viewed as a 
mixing of radiation with different thermodynamic temperatures. 
As a consequence, the Plankian radiation spectrum undergoes a
$y$-type distortion: it may be characterized by a Compton-$y$
parameter.  Damping of perturbations that `enter' the horizon
at epochs $z < z_{th}$ when thermalization processes had ceased to be
effective, but at epochs $z > z_{c} \sim 1.6 \times 10^{5}$
when Compton scattering was capable of driving the radiation
spectrum to kinetic equilibrium, resulted in a transformation of the
$y$ distortions to $\mu$-type distortions characterised by a
chemical potential (Daly R.A. 1991; Hu, Scott \& Silk 1994).
These $\mu$ distortions are {\it inevitable} in `standard' theories of
structure formation; the magnitude of the distortion is a probe of the
amplitude of the primordial perturbation spectrum at large $k$ modes 
or --- combined with the COBE normalization for small $k$ modes ---
may be viewed as a probe of the index $n$ of a power-law form spectrum.

Besides the inevitable distortions from structure formation, any processes
that release radiant energy in the redshift interval $z_{th} > z > z_{c}$
would result in a $\mu$ distortion in the CMB today.  For example, the
decay of particles with half lives in this range of cosmic epochs could be
probed via the expected $\mu$ distortions in the relict CMB (Silk \&
Stebbins 1983).

A simple $\mu$ distortion in a Plankian spectrum manifests as a divergence
in the thermodynamic temperature with increasing wavelength.  However,
because the thermal bremsstrahlung, which is one of the processes
responsible for the thermalization
of the radiation at redshifts $z > z_{th}$, has a frequency dependence and
is more effective at longer wavelengths, distortions 
at long wavelengths are thermalized and consequently the thermodynamic
temperature of the radiation may be expected to have an extremum in its
deviation.  Maximum distortion occurs at about a wavelength
$\lambda_{m} \sim 30$~cm (Burigana, Danese \& De Zotti 1991).

$\mu$ distortions in the CMB that arise as a consequence
of processes at epochs $z_{th} > z > z_{c}$  may, 
therefore, be best constrained by measurements of the absolute temperature
of the CMB at frequencies about 1~GHz.  We have attempted to make a
measurement of the thermodynamic temperature of the CMB at 1280~MHz.

\section{The receiver system}

The receiver used for the measurement described in this
work is a modified form of the L-band front-end package built for
the Giant Metrewave Radio Telescope (GMRT; Swarup et al. 1991, 1997).  
This receiver was designed
to be a package mounted at the prime focus of the GMRT antennae and its
feed horn was designed to illuminate the parabolic dish aperture.
For our experiment, the receiver package was placed on the ground and
pointed at zenith.  A schematic of the receiver configuration
is shown in Fig.~1.

The corrugated feed-horn used in the GMRT receiver was replaced by
a flat plate with concentric corrugations; this is directly
connected to the top of the quadridged
orthomode transducer (OMT) and serves as the interface between
free space and the transducer.  The OMT is a circular cylindrical
waveguide and at the bottom end the signal
is transduced onto co-axial cables.  Although the OMT provides dual
orthogonal polarizations, only a single linearly polarized
signal (`V' port signal) has been used for the background measurement.
This co-axial signal is fed to a
low-noise amplifier (LNA) through a circulator.  In section~3 we ellaborate
on issues related to the introduction of this circulator.
The LNA output is connected to a chain of 
band-pass filters and is  effectively limited to a 120~MHz band
centred at 1280~MHz. The receiver package is at ambient temperature
and no part is cooled.

The receiver output is connected to 
a spectrum analyser through a commercial Miteq amplifier. 
The noise figure of the spectrum
analyser is so high that in the absence of the Miteq amplifier, the 
spectrum analyser's noise contribution degrades the 
signal-to-noise by contributing
significantly to the overall system temperature.
The power spectrum is measured by reading the trace of the
spectrum  analyser using a general purpose interface board (GPIB)  
and the spectral data is acquired in a personal computer. 

The receiver and feed assembly are 
surrounded by an aluminium shield to minimize
ground radiation leakage into the feed. The design of the shield and
measurements of the ground contribution are detailed 
respectively in sections~4 and 12.

\begin{figure}
\epsfig{file=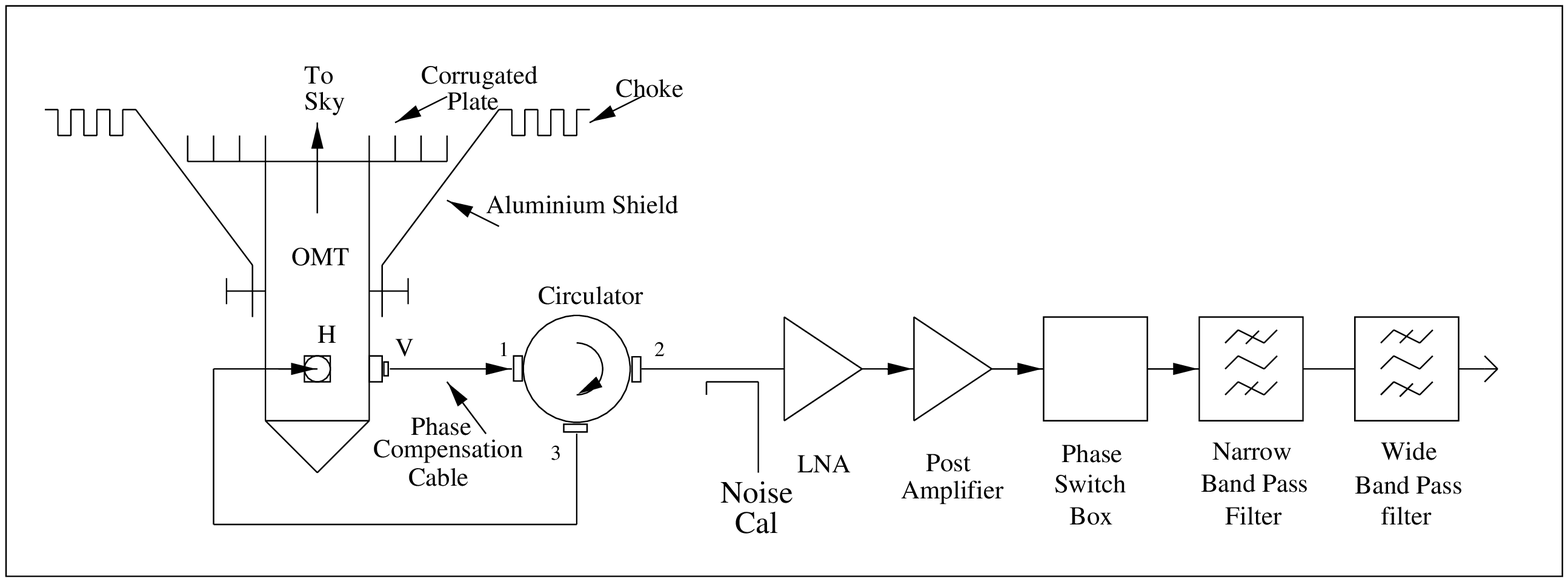,height=8cm,width=13cm,angle=0}
\caption[]{Schematic diagram of the 1280~MHz receiver}
\end{figure}     
   
The cosmic microwave background  temperature $(T_{CMB})$ 
has been measured  using the above
receiver at 1280~MHz over a bandwidth of 100~MHz.  The signal power incident
at the corrugated plate (feed) will be an additive combination
of contributions from
the cosmic microwave background, emission from the Milky Way
Galaxy, atmospheric emission  and ground contribution through 
sidelobes of the feed assembly.  
Therefore, the noise temperature entering the feed plate 
will be the sum of the 
Galactic temperature ${T_{Gal}}$,
atmospheric temperature $T_{atm}$, ground temperature ${T_{gnd}}$ 
and cosmic
microwave background temperature $T_{CMB}$.  For an accurate 
determination of
$T_{CMB}$, it is necessary to know the values of $T_{Gal}$, $T_{atm}$, 
and $T_{gnd}$
so that these may be subtracted from the measurement of power at the
feed plate.  In sections below, these contributions are separately estimated.

The noise power entering the feed propagates via the OMT and circulator
before amplification in the LNA.  Beyond this point, the high gain in the
LNA and the negligible signal-to-noise-ratio degradation in successive stages
ensure that any additive noise contributions are insignificant.
Following the calibration scheme discussed below in section~7,
the power measured by the spectrum analyser is calibrated to
represent the equivalent noise power referred to the input port (port~1)
of the circulator.  Expressed in Kelvin, this noise power represents
the total system temperature $T_{sys}$.  
In order to estimate the noise power entering the feed and isolate
this component in a measurement of the total system temperature,
it is necessary to separately measure system parameters like the receiver 
temperature ${(T_R)}$ and the reflection coefficient $(\Gamma)$ and  
the absorption coefficient $({\alpha})$ of 
the feed assembly (corrugated plate + OMT).  

We now present a formulation of the measurement problem.
Let ${{T_a}^{\prime\prime}}$ represent the total external
signal  incident on  the corrugated plate (feed):  

\begin{equation}
{T_a}^{\prime\prime} = T_{Gal} +T_{atm}+T_{gnd}+T_{CMB}.
\end{equation}

\noindent The total (external + internally generated) power  at
the input terminal (port~1) of the circulator is given by  

\begin{equation}
T_a = {{T_a}^{\prime\prime}}(1-\Gamma^{2})(1-\alpha) + 
					\alpha T_{amb},
\end{equation}

\noindent where the antenna temperature $T_a$ is represented as an
additive sum of the external power --- corrected for attenuations and
reflections in the feed assembly --- and the internal noise generated
as a consequence of the loss in the feed assembly.  The feed assembly is
at ambient temperature $T_{amb}$.
The system temperature $T_{sys}$, as referred to the
input of the circulator (port 1), is given by 

\begin{equation}
T_{sys} = {{T_a}^{\prime\prime}}(1-\Gamma^{2})(1-\alpha) + \alpha T_{amb} 
		+ T_R. 
\end{equation}

\noindent  In this  equation, the first term on the right
represents the net external signal 
present at the input of the circulator, 
the second term represents the thermal noise 
contribution from the  feed assembly   
and the third term is the receiver temperature
as referred to the circulator input. 
Measurements of the various parameters in the above equations  
are necessary for an estimation of 
the absolute value of $T_{CMB}$ from a measurement of the 
calibrated system temperature at the input of the circulator.

Photographs of the system, as used for the CMB measurement, are shown in
Figs.~2 and 3.  The receiver electronics is inside a 
(40~cm x 60~cm x 100~cm)
rectangular box.  Fig.~3 shows the system with the shield lowered: visible
at the top of the receiver box is the corrugated plate --- which serves as
the feed --- and this is connected to the OMT 
housed inside the box.  As seen in the figure, the shields are constructed
from trapezoidal aluminium plates and when the shield is mounted as 
shown in Fig.~2, 
the top of the receiver box, including the feed plate,
is shielded on all four sides and the shield minimizes ground 
radiation entering the system.  
The receiver output is accessible
below the shield and is connected to the 
spectrum analyser, mounted separately on a trolley, via an RF co-axial cable.  
The trace data
obtained in the spectrum analyser is recorded via a GPIB interface 
in a computer.  

\begin{figure}
\epsfig{file=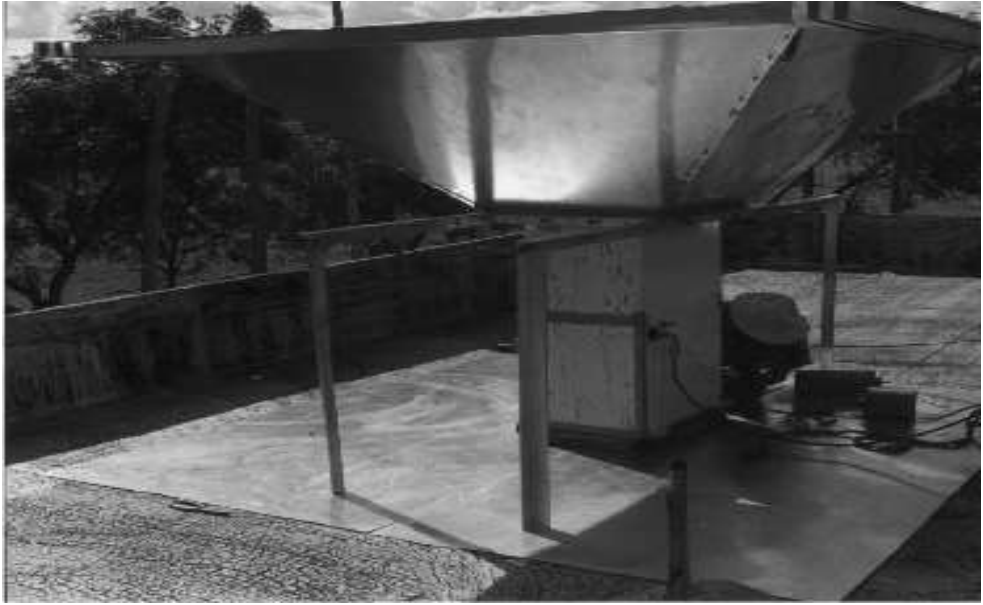,height=13cm,width=8cm,angle=-90}
\caption[]{Photograph of the 1280~MHz receiver with the aluminium shield
in place}
\end{figure}     

\begin{figure}
\epsfig{file=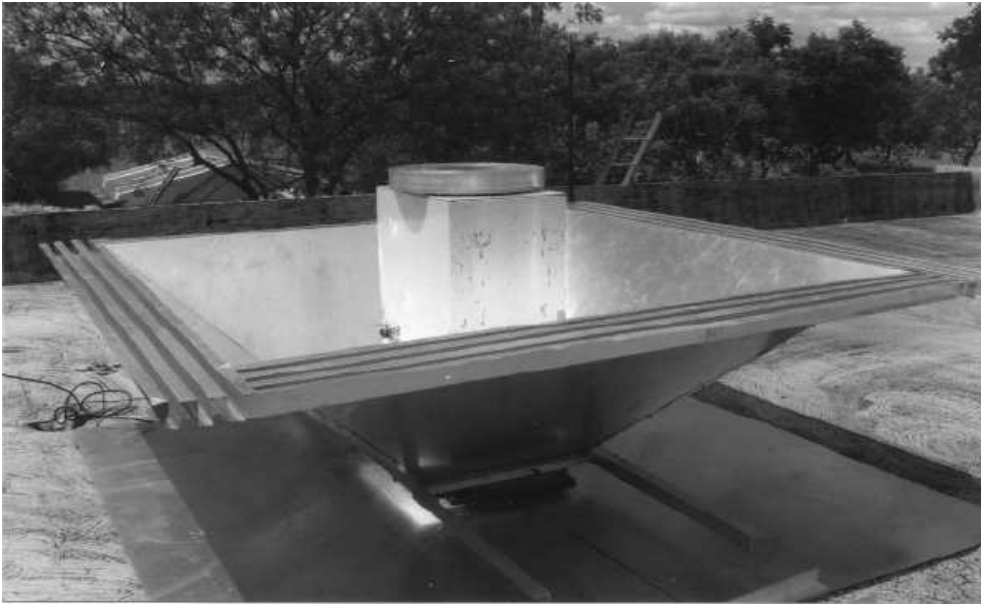,height=8cm,width=13cm,angle=0}
\caption[]{Photograph of the 1280~MHz receiver with the aluminium shield
lowered to the ground}
\end{figure}     

\section {Noise reduction techniques}

The circulator has been introduced at the input of the 
LNA in order to isolate the LNA from reflections
in the corrugated plate-OMT assembly while observing the sky. 
Multiple reflections, if allowed, may cause the LNA noise temperature
to be significantly dependent on the input matching and, consequently,
the LNA temperature as measured under test conditions in the laboratory
(section~7) may be different from the LNA temperature when the 
receiver is configured to observe the CMB sky temperature.

As shown in Fig.~1, the V-channel output from the OMT, which represents
the sky signal $T_a$ that is to be measured, is connected to port~1
of the circulator and propagates, with a small loss, on to the LNA
via port~2 of the circulator.  The `cold load' at port~3 of the
circulator is derived from the `H' port sky signal of the OMT.
In addition, the `H' port also serves as a conduit for radiating the
LNA noise power --- which propagates from the LNA into port~2 of the
circulator --- to the sky.  We have verified that the isolation between
the `V' and `H' ports exceeds 30~dB.

Any noise power (sky signal or reflected LNA noise) 
that enters port~3 of the circulator from the `H' port of the OMT
would propagate to the LNA as a spurious additive power via two paths:
(i) leakage directly from port~3 to port~2 and (ii) by being reflected
from the `V' port of the OMT.  We have selected the length of
the cable connecting the `V' port of the OMT to port~1 of the circulator
so that the two signal paths would have a difference of half-integral
wavelengths.  Consequently, the `phase-compensation' cable (Fig.~1) 
causes a partial cancellation of the spurious signal power and enhances the
isolation, by an additional 5~dB, of the 
LNA from spurious signals entering via port~3.

\section {Ground shields}

The OMT in the receiver is a uniform circular waveguide.  Usually, the
free-space impedance is matched to the OMT impedance via a feed; in
the case of the GMRT, a wide-flare-angle corrugated scalar horn
is used.  Derivation of the sky temperature from system temperature
measurements requires a characterization of the losses in the
feed assembly.  The losses in the OMT
are measured, as discussed below, by taking advantage of the fact
that the OMT is uniform; however, the losses in any tapered horn
are difficult to measure.  Therefore,
we omitted the feed horn from our receiver system. As
a consequence: (i) on the positive side we do not have to measure
any losses in any feed system, (ii) the negative aspect is that 
the OMT is now poorly matched to free space.  The mismatch is
measured, as described below, by taking advantage of 
reciprocity in the passive OMT.

Steps are taken to minimise the ground 
contribution as much as possible. We have chosen to
have a simple corrugated plate on the top of the OMT to reduce
the far-off sidelobes in the instrument response. The plate reactively,
rather than resistively, attenuates any incoming power from directions
at large angles to the OMT axis.  Because this interface between the
OMT and free space, which replaces the feed horn, acts only on signals
incident at large off-axis angles where the OMT response is low, and
because the corrugated plate is a reactive block, we do not expect
any significant contribution from the plate to the system temperature.

We have also constructed a solid aluminium ground shield to further
reduce the ground contribution.  In order to reduce
diffractive leakage of ground emission around the edges of the shield, 
chokes are placed all along the periphery 
of the shield around the receiver.  In addition, 
as shown in Figs.~2 and 3, an aluminium
sheet covers the ground directly below the system.

\begin{center}
\begin{figure}
\epsfig{file= 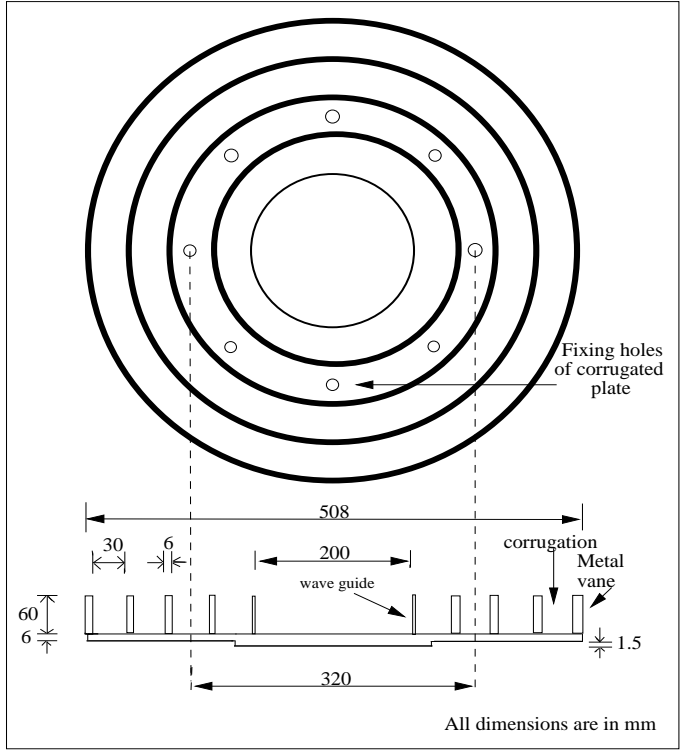,width=9 cm,height=10 cm,angle=0}
\caption[]{Schematic diagram of the corrugated plate}
\end{figure}     
\end{center}

\subsection{Design of the corrugated plate}

A circular aluminium plate of 6~mm thickness and with 
outside diameter equal to
508~mm and inner diameter equal to 200~mm is used as a base for the 
choke slots.  Three concentric 
corrugations are formed on this plate using 
6~mm thick and 60~mm wide aluminium strips bent in the form of 
rings of different diameters 
as shown in the Fig~4.  The width of the circular strips are chosen to be
$\lambda/4$ at the centre frequency (1280~MHz); this makes
the corrugations $\lambda/4$ deep.
The rings are spaced  at approximately $ 0.125 \lambda$ (30~mm) 
apart and this dimension  
will be the width of the concentric corrugation slots.
These circular rings 
are fastened onto the metal plate using screws.
For good electrical continuity, tinned copper 
braid is introduced between the circular
rings and the base plate before fastening.  The 
corrugated plate is attached to the OMT via
an interface plate.  The radiation pattern of 
the OMT with the corrugated plate is shown in Fig.~5.
It is clear from the figure that the radiation 
patterns in both V and H planes are 
symmetrical and have a 3~dB beam width of $\approx 60^{\circ}$.
The introduction of the plate marginally reduces the 3~dB width
of the main lobe; as compared to the radiation pattern of the
OMT (without the corrugated plate), the feed assembly has significantly
($> 10$~dB) reduced sidelobes.

\begin{figure}
\epsfig{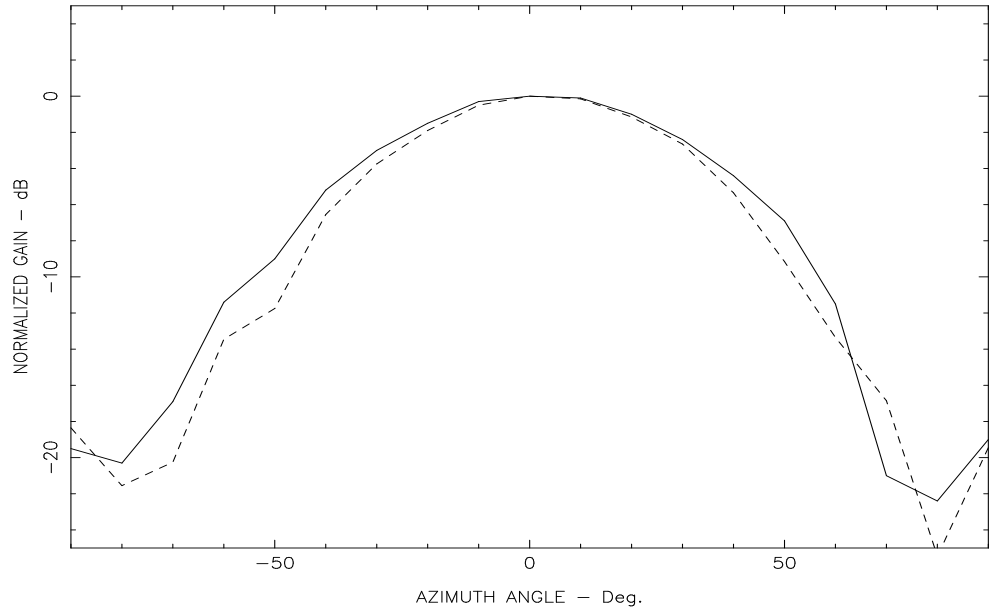}
\caption[]{The radiation pattern of the modified feed assembly in the E
(solid line) and H (dashed line) planes at 1280 MHz}
\end{figure}     

\subsection{Design of the ground shield}

An aluminium shield is used  
to reduce the ground contribution to the system 
temperature of the receiver.  The shield is made 
of four trapezoidal parts of similar dimensions: 
they all have a short vertical straight section of about 10~cm 
height followed by a 1~m long section bent at an angle of
$45^{\circ}$ to the vertical axis.  At the top of the $45^{\circ}$ section
is attached a choke formed of thin aluminium sheet.
The purpose of the choke is to reduce the 
ground noise entering into the system over the top of the shield through 
edge diffraction.  A schematic of the shield, along with the dimensions, 
is in Fig.~6. The shield is fixed to an
aluminium square base frame made using 1.5-inch `L' sections.
The frame may be fixed to the receiver at any height. Too high a position
disturbs the radiation pattern --- as measured by a change in the return-loss
of the feed assembly --- whereas too low a position allows ground pick-up;
the final position was found not to be critial as long as 
the location was not extreme.

\begin{figure}
\epsfig{file=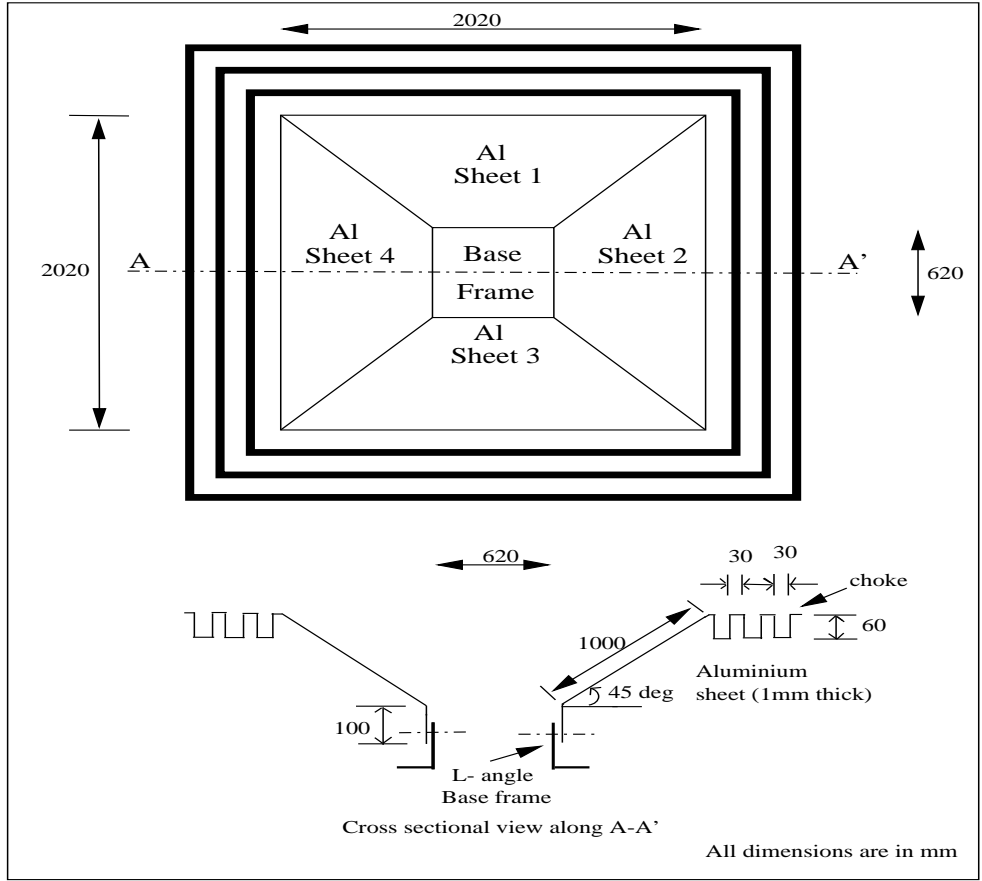,height=12cm,width=13.0cm,angle=0}
\caption[]{Schematic diagram of the Aluminium shield}
\end{figure}

\section{The absolute temperature calibration}

The absolute temperature calibration of the receiver was performed
using resistances immersed in liquid baths whose physical temperature
was measured using a platinum-wire resistance thermometer.
In the laboratory, the reference baths were used to 
determine the noise temperature of the
calibration signals injected at the LNA input.  In the field, the
noise diodes provided the secondary reference.

The platinum resistance thermometer was built using a Pt-100 sensor
and a constant current source.  This 
thermometer was calibrated by immersing it in the bath of a
commercial temperature controlled 
chiller (Ultra Temp 2000) and examining the variation in the voltage
across the platinum resistance with the bath temperature.
A precision mercury thermometer was
placed in the bath in close proximity to 
the platinum resistor to measure the bath temperature and 
a calibration relationship was determined for the platinum resistance.
The Pt-100 sensor was also
calibrated at liquid nitrogen temperatures by placing it in a liquid nitrogen
bath whose temperature was measured using (a) a 
calibrated cryogenic thermometer
in the Indian Space Research Organization (ISRO) at Bangalore, India and (b) a
Pt-1k commercial calibrated sensor from Rosemount Corporation. 
The relations obtained from the calibration measurements for 
the determination of temperature using the measured voltages were 

\begin{equation}
    V = 195.531799 + 0.757149 (T+273.15), {\rm ~~and}  
\end{equation}

\begin{equation}
    V =  0.83 T - 25.88,
\end{equation}

\noindent where $T$ is the temperature of the sensor in Kelvin and 
$V$  is the voltage measured across the sensor in Volts.
Equation~4  is used while measuring the temperature in the 
range 0--100$^{\circ}$~C and equation~5 is used in the temperature
range 50--100~K.  The calibrated platinum resistance thermometer
is estimated to give the temperature of liquid nitrogen baths 
accurate to 
within 0.2~K and the temperature of water in the range 0--100$^{\circ}$C
to within 0.1~K.

\begin{figure}
\epsfig{file=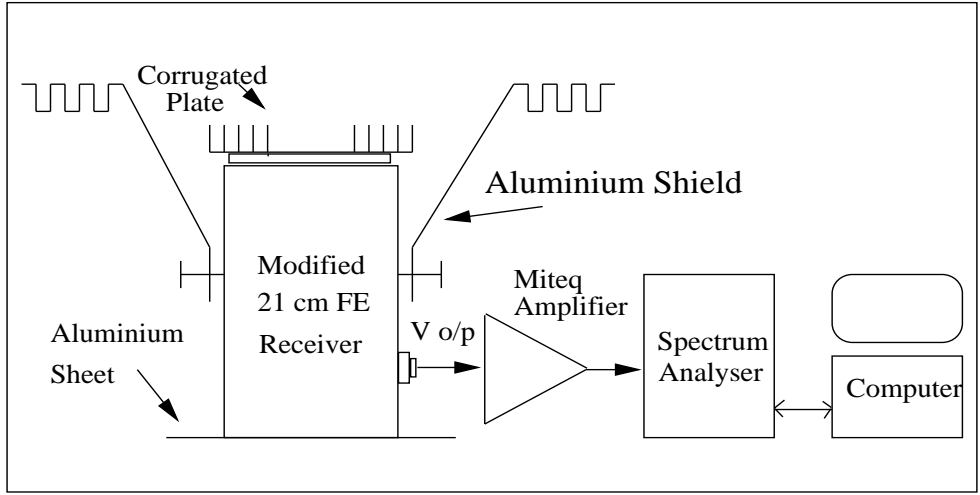,height=8cm,width=13.0cm,angle=0}
\caption[]{The experimental setup for the acquisition of the data}
\end{figure}     

\section{The data acqusition system}

The setup for the measurement of the 
cosmic microwave background temperature is shown 
in Fig.~7. The sky radiation collected  by the corrugated plate-OMT 
is amplified and band limited within the 
receiver. The V-channel output of the receiver is connected to the spectrum 
analyser to measure the 
power received.  Using a personal computer (PC) interfaced to the spectrum
analyser, the instrument is initialized  and
trace data are acquired from the spectrum analyser.
The spectrum analyser is configured to 
cover a 100-MHz span around 1280-MHz center frequency with a 
1-MHz resolution bandwidth. The video-averaging function in
the spectrum analyser is disabled.

A stable noise may be additively injected into the signal path close to 
the input of the LNA.
This noise CAL is periodically switched ON and OFF 
during the acquisition and serves as a reference power for
calibrating the receiver: the CAL remains ON for a second and 
OFF for a second.
The appropriate CAL power level and filter selections are manually set
prior to data acquisition. 

The spectrum analyser is interfaced to the PC via a GPIB and
the recording of the power spectrum is made by acquiring the
trace data, as displayed by the spectrum analyser, into the PC.  
During each readout, a 401-point
frequency spectrum is read from the spectrum analyser. 
In each 2-sec period, 120~x~2 traces are read (120 with CAL in ON state
and 120 with CAL OFF) and the traces are averaged separately for
the CAL ON and CAL OFF states over the 100~MHz band. 
The laboratory determination of the CAL noise 
temperature ($T_{cal}$)--- this
procedure is described in section~7 --- is used to
calibrate the temperature scale of the acquired trace data.
It may
be noted here that the internal calibration of the spectrum analyser
is not used for determining the temperature scale.

This acquisition system was used not only for making measurements
of the CMB sky power but also in all the calibration measurements
described below for determining $T_R$, $T_{cal}$
and $\alpha$.  This ensured that all calibration measurements were
made over the same frequency band and with the same spectral weighting.

Because we use the spectrum analyser to acquire the data, which 
effectively sweeps a narrow filter with 1~MHz width 
across the 100~MHz band in order to
measure the spectrum, the effective integration time is only 
about 1 per cent of the observing duration.  With our system,
spectra are measured with a fractional accuracy 
of about 0.13 per cent
if data are acquired for a period of one hour.  

\section{Laboratory measurement of $T_R$ and $T_{cal}$}

$T_{R}$  and $T_{cal}$ are the 
receiver noise temperature and calibration noise temperature 
referred to the input port of the circulator (port 1). 
The measurements of $T_R$ and $T_{cal}$ were made 
by comparing their noise powers with those from 
resistor termiations placed in
standard temperature baths containing separately liquid 
nitrogen and ambient-temperature water.

For this measurement, as shown in Fig.~8,
port~1 and port~3 of the circulator are terminated in 50~$\Omega$
loads.  Port~2 is connected to the LNA input and the signal continues
through to the rest of the receiver and acquisition system.
The termination at port~3 is kept 
immersed in liquid nitrogen, while the termination 
at port~1 is separately immersed in liquid 
nitrogen and then ambient temperature water for the process 
of calibration.  The calibrated platinum
resistance thermometer is used to measure the temperature of 
the 50~$\Omega$ load at port 1. The calibration signal $(T_{cal})$
is injected in alternate seconds 
of time as would be done during the sky temperature measurement.  

The measured uncalibrated noise powers $P_{on}$ and $P_{off}$, 
corresponding to the states when the CAL is ON and OFF, are used to
compute $y$-factors: $(P_{on}-P_{off})/P_{off}$.  The
$y$-factors $x$ and $y$ corresponding to the measurements with the
port~1 load at ambient temperature $T_{amb}$ and at liquid nitrogen
temperature $T_{N2}$ are used to estimate the LNA noise temperature

\begin{equation}
T_{R} = {(x T_{amb} - y T_{N2}) \over(y-x)}  
\end {equation}

\noindent and the CAL noise temperature

\begin {equation} 
T_{cal} = x(T_{R}+T_{amb}) .     
\end {equation}

\begin{figure}
\epsfig{file=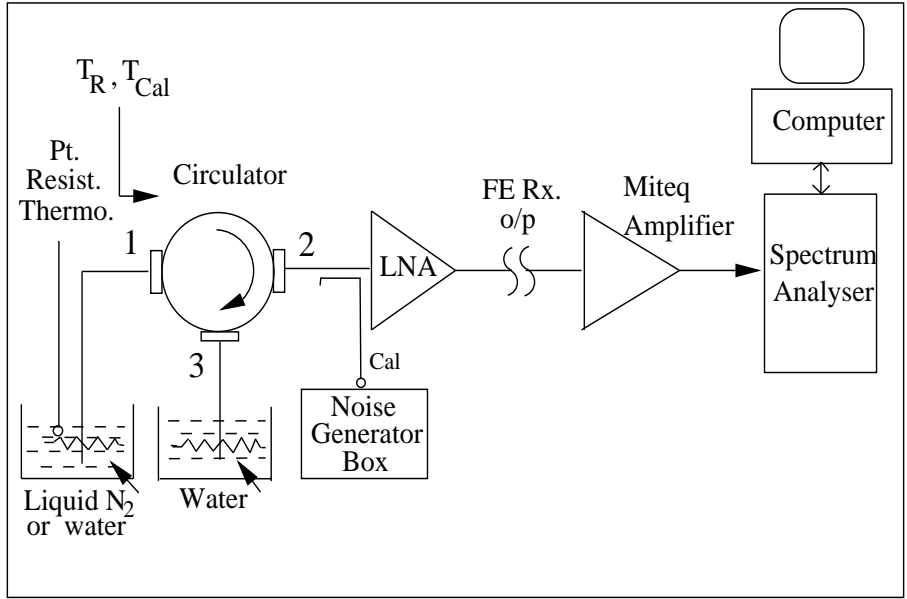,height=8cm,width=12.0cm,angle=0}
\caption[]{The experimental setup for the measurement of Receiver
temperature ${T_R}$ and Cal temperature $T_{cal}$}
\end{figure}     
   
The 1-$\sigma$ errors in the $T_R$ and $T_{cal}$ estimates 
owing to measurement noise (corresponding to the finite bandwidth and
averaging time of the measurement) are 0.38~K and 0.20~K respectively. 
The errors due to the uncertainty in the temperatures of the liquid
nitrogen and ambient temperature water baths are 0.34~K for $T_R$ and
0.08~K for $T_{CAL}$.
The measured values of $T_{R}$ and ${T_{cal}}$ are

\begin{equation}
{T_{R}}   = 52.12 \pm 0.51~{\rm K~~~and} 
\end{equation}

\begin{equation}
{T_{CAL}} = 70.166 \pm 0.215~{\rm K}, 
\end{equation}

\noindent where the uncertainties quoted represent 1-$\sigma$ errors.

\section{The measurement of the reflection coefficient $\Gamma$}

\begin{figure}
\epsfig{file=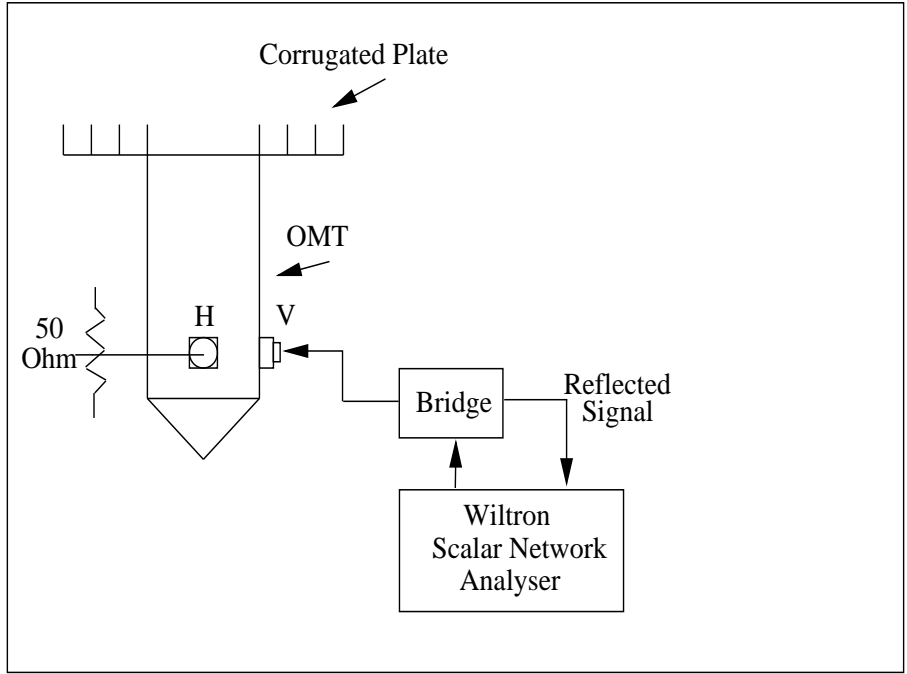,height=9cm,width=12.0cm,angle=0}
\caption[]{The experimental setup for the measurement of  the reflection
co-efficient of the feed assembly}
\end{figure}     

Due to the impedance mismatch between the feed assembly and the sky, 
part of the incident sky power is reflected back to the sky.  The 
fractional power reflected is characterized by the voltage reflection 
coefficient ($\Gamma$) of the feed assembly.  
Because the feed system is passive and hence reciprocal, the system may be
operated as a radiator in order to measure the reflection coefficient.
The setup (Fig.~9) consists of a scalar network
analyser connected to the V-channel of the feed assembly; the H-port of
the OMT is terminated at a 50~$\Omega$ load for this measurement.
The instrument injects power into the OMT port and measures the
return loss: the fraction that returns as a reflected power.
 
The reflection coefficient of the feed assembly, as computed  
from the return loss measurement, is

\begin{equation} 
    \Gamma = 0.1265 \pm 0.01~. 
\end{equation}

\noindent The uncertainty in this estimate has been assumed to be 
the measurement accuracy of the scalar network analyser.

\section{The measurement of the absorption coefficient $\alpha$}

The absorption coefficient $\alpha$ includes contributions from
the OMT and the phase-compensation cable (and the associated connector).
For reasons stated above, it has been assumed that the ohmic loss in 
the corrugated plate is negligible.

Advantage is taken of the cylindrical shape of the OMT:
$\alpha$ for a single OMT along with its phase compensation cable 
is determined by combining an identical pair 
back-to-back and measuring the ohmic loss in the combined 
system. The apertures of the two OMTs are bolted together, the H-channel
ports are terminated in 50~$\Omega$ loads and the V-channel ports are
connected to identical-length phase-compensation cables. 
The procedure described in section~7 for the measurement of the
receiver temperature $T_R$ and the calibration noise temparture
$T_{cal}$ is repeated with the back-to-back 
OMT pair introduced between the port~1 of the 
circulator and the load termination that is separately placed in
liquid nitrogen and ambient water baths.  The measurement configuration
is shown in Fig.~10.

\begin{figure}
\epsfig{file=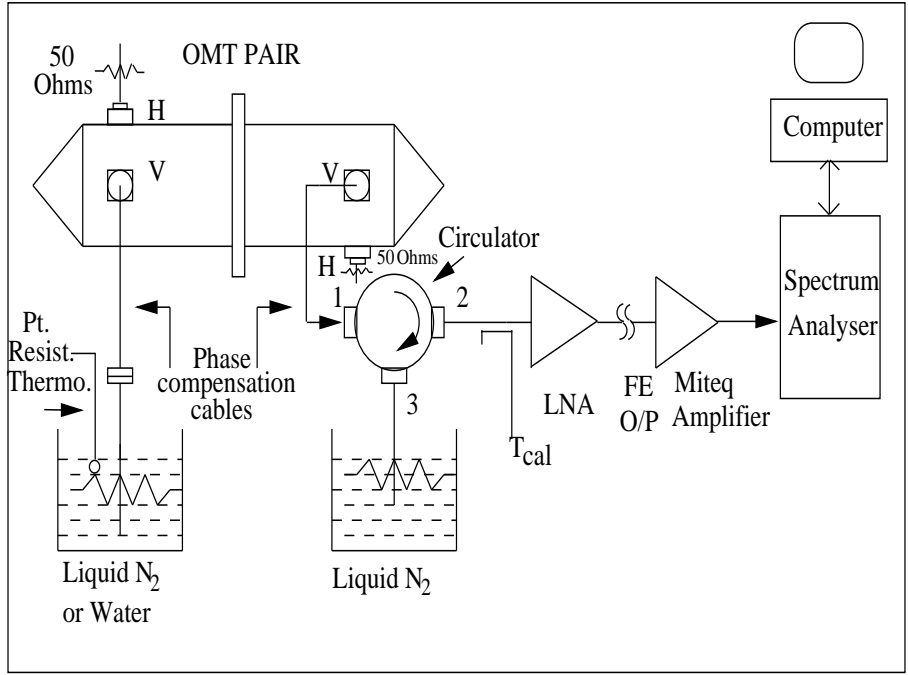,height=9cm,width=12.0cm,angle=0}
\caption[]{The experimental setup for the measurement of 
the absorption co-efficient of an OMT pair}
\end{figure}     

In this configuration, the measured receiver and calibration noise
temperatures are, respectively, 

\begin{equation}
  {T_R}^{\prime} = {\left[{\alpha_{p} T_{amb}\over 
(1-\alpha_{p})(1-{\Gamma_{o}^{2}\over 4})} + {{T_R+T_L} \over (1-\alpha_{p})
{{(1-{\Gamma_{o}^2\over 4})}^2}} \right]}  
\end{equation}
 
\noindent and

\begin{equation}
  {T_{CAL}}^{\prime} = \left[{T_{CAL} \over 
	{(1-\alpha_{p}){{(1-{{\Gamma_{o}^2}\over 4})}^2}}}\right]  
\end{equation}

\noindent in terms of the $T_R$ and $T_{cal}$ values determined in section~7.
In these equations, $\alpha_{p}$ is the absorption coefficient of the OMT pair 
along with the two phase compensation cables,
$\Gamma_{o}$ is the voltage reflection coefficient of the OMT pair and
$T_L$ is the small leakage signal at port~2 due to the load at port~3 of the 
circulator.  

From the measurements, we find that the reflection coefficient of each OMT
in this configuration is $\Gamma_{o}/2 = 0.0912$ and 
the absorption coefficient was determined to be

\begin{equation}
  \alpha_{p}/2 = 0.031059 \pm 0.0016.                               
\end{equation}
 
\section{The Galactic contribution $T_{Gal}$}

We estimate 
the Galactic contribution to the sky signal from 
a 408~MHz all sky map (Haslam et al. 1981).  
The sky image is smoothed to the resolution of the feed assembly and
the brightness temperature at 1280~MHz is computed assuming a spectral
index of $-2.7$ for the temperature spectrum.

The sky background observations were made at zenith
from the radio observatory site at Gauribidanur which is at a
latitude of $+13^{\circ}$\llap.5.  
The observations were made at night to avoid 
the Sun. The sky region observed was away from the Galactic plane
and was at approximately RA:10--12$^h$ and DEC:+13$^{\circ}$\llap.5.
The value of the sky brightness $T_{408}$ 
towards  this region is $18 \pm 2$~K in the 408~MHz all sky map.  Assuming 
that the CMB temperature at 408~MHz is 2.7~K, the Galactic background 
temperature at 408~MHz may be $15.3 \pm 2$~K.  
The Galactic contribution at 1280~MHz is estimated to be

\begin{equation}
     T_{Gal} = 0.9 \pm 0.3~{\rm K}. 
\end{equation}

\noindent The sources of uncertainty in this estimate are the uncertainty
in the 408-MHz absolute temperature scale and the uncertainty in the
spectral index of the Galactic background.

\section{The atmospheric contribution $T_{atm}$}

\noindent A multiple slab model for the atmosphere (developed by 
J. Cernicharo and M. Bremer at IRAM)
has been used to estimate the contribution of the atmosphere to the 
total system temperature.  The model relates the physical parameters of the 
atmosphere slabs to be consistent with the altitude and latitude
of the site and the local temperature, pressure, and zenith column 
density of water vapour. A prediction is made for the atmospheric
emission brightness temperature.  
The Gauribidanur observatory is at 
13$^{\circ}$36$^{\prime}$16$^{\prime\prime}$ North latitude and at an 
elevation of 686~m; the ground temperature of the atmosphere was
25$^{\circ}$C, pressure 931~mB and the zenith column 
density of water vapour was $\le 10$~mm 
during the sky temperature observations.
Using the model, the atmospheric contribution $T_{atm}$ at zenith is  
estimated to be 

\begin{equation}
 T_{atm} = 1.55\pm 0.2~~{\rm K}.  
\end{equation}

\section{The ground contribution $T_{gnd}$}

\noindent The ground contribution is measured by radiating a 1280~MHz 
signal out through the OMT pointed at zenith (with the shield in place)
and measuring the radiated signal with a dipole.  
This measurement of radiated power towards the ground yields the
isolation between the ground and receiver and, invoking reciprocity, 
it is estimated that less than about
0.1\% of the ground noise temperature enters the OMT.
To avoid the leakage via gaps between the bottom of the shield and the
receiver box, the ground close to the receiver
was covered with aluminium sheets (see Fig.~2).  
With these ground sheets, ground shield and the
corrugated plate all in place, and   
assuming the temperature of the ground to be 30$^{\circ}$C, 
the ground contribution is estimated to be

\begin{equation}
T_{gnd} = 0.3\pm 0.3~~{\rm K}.
\end{equation}

\section{The measurement of the CMB brightness temperature}

\noindent The measurement of the sky brightness temperature 
was done from the Gauribidanur observatory site towards zenith.
The observations were made at a time when 
the Galactic plane was far from the primary beam of the `telescope'
and at night to avoid the Sun: the observations were carried
out between $10^h$ and $12^h$ LST (i.e. 10 pm and 12 pm local time) during 
which the zenith sky was 
at RA:10-12$^h$ and DEC:+13$^{\circ}$\llap.5.
 
\begin{table}
\epsfig{file=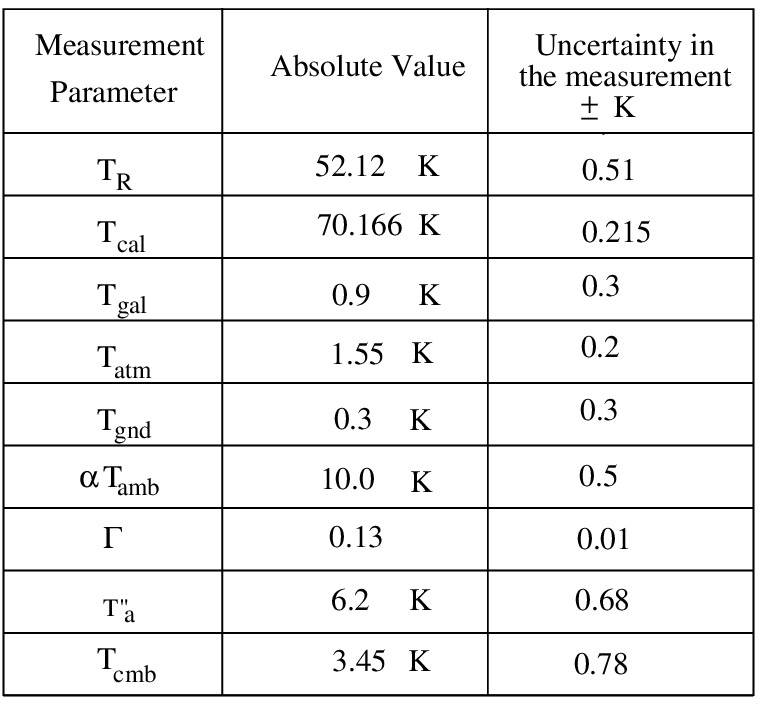,height=8cm,width=13.0cm,angle=0}
\caption[]{Values of various noise temperatures
           measured}
\end{table}     
In Table~1 we have gathered all the calibration measurements
from which we have derived the 	
the sky temperature $T_{a}^{\prime\prime}$ 
to be $6.2\pm0.68$~K.  From equation~1,
the brightness temperature of the cosmic microwave background  
$T_{CMB}$ was estimated to be 

\begin{equation}
T_{CMB} = 3.45\pm0.78~{\rm K}. 
\end{equation}	   
 
\section{A comparison with some earlier measurements}
 
\begin{table}
\epsfig{file=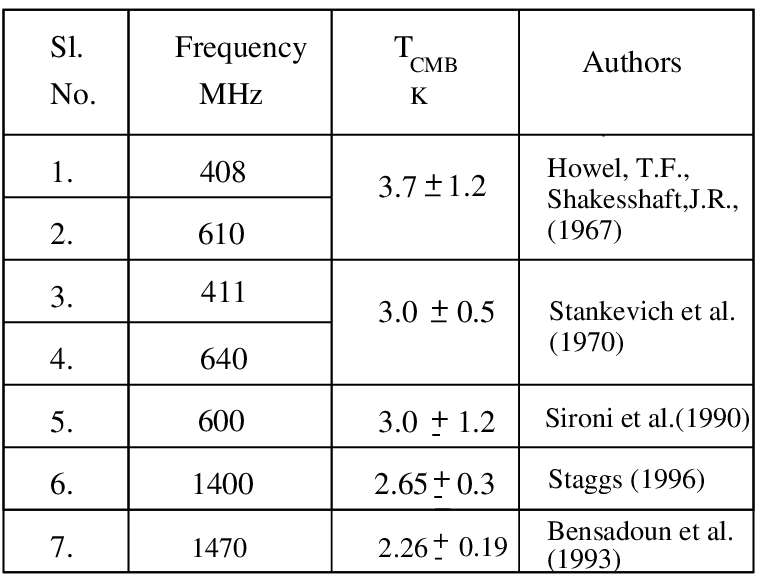,height=8cm,width=13.0cm,angle=0}
\caption[]{ Some earlier measurements of $T_{cmb}$ at 
frequencies below 1.5~GHz}
\end{table}  
    
Our measurement is consistent, within the errors, with the $COBE$-$FIRAS$
measurement. 

Previous measurements of the absolute temperature of the CMB
at frequencies below 1.5~GHz are listed in Table~2.
Below our frequency of 1280~MHz, the closest measurement is that of
Sironi et al. (1990) at 600~MHz: their measurement is 
consistent --- within their 1-$\sigma$ error --- with the $COBE$-$FIRAS$
value; however, it may be noted that our measurements have a 
greater precision.  At frequencies above 1280~MHz, the measurement
by Staggs (1996) at 1400~MHz has a precision exceeding our 
measurement and is also consistent (within the 1-$\sigma$ errors) 
with the $COBE$-$FIRAS$ value.  Our measurement at 1280~MHz, as well as
these measurements at 600 and 1400~MHz, do not indicate any distortions
in the CMB spectrum and are consistent with $\mu = 0$.  

It may be noted, 
however, that the measurements of Levin et al. (1988) at 1410 MHz and
those of Bensadoun et al. (1993) at 1470~MHz both imply lower
thermodynamic temperatures for the CMB at these frequencies: their
measurements of $2.11 \pm 0.38$~K and $2.27 \pm 0.25$~K are about 
1.6--1.8-$\sigma$ below the $COBE$-$FIRAS$ value. Our
measurement at 1280~MHz is inconsistant with such a low value 
for $T_{CMB}$ at the 1.6-$\sigma$ level.

\section {Acknowledgements}

\noindent We thank V. Radhakrishnan, A. A. Deshpande and N. Udaya Shankar for 
stimulating discussions had during the course of this work.

\label{lastpage}

\end{document}